\documentclass[10pt,superscriptaddress,twocolumn,amsmath,amssymb,aps,prb]{revtex4-1}
\usepackage{mathrsfs}
\usepackage{graphicx}
\usepackage{dcolumn}
\usepackage{bm}
\usepackage{color}

\renewcommand{\Im}{\operatorname{Im}}

\newcommand{\be}{\begin{equation}}
\newcommand{\ee}{\end{equation}}
\newcommand{\bea}{\begin{eqnarray}}
\newcommand{\eea}{\end{eqnarray}}

\begin{document}
\title{Quantum Chaos of the Bose-Fermi Kondo model at the intermediate temperature}
\author{Xinloong Han}\email{hanxinloong@gmail.com}
\affiliation{Department of Physics and Center of Theoretical and Computational Physics, The University of Hong Kong, Hong Kong, China}
\author{Zuodong Yu}
\affiliation{Zhejiang Institute of Modern Physics and Department of Physics, Zhejiang University, HangZhou, 310027, China}
\affiliation{Zhejiang Province Key Laboratory of Quantum Technology and Device, Zhejiang University, HangZhou, 310027, China}

\date{\today}
\begin{abstract}
We study the quantum chaos in the Bose-Fermi Kondo model in which the impurity spin interacts with conduction electrons and a bosonic bath at the intermediate temperature in the large $N$ limit. The out-of-time-ordered correlator is calculated based on the Bethe-Salpeter equation and the Lyapunov exponent $\lambda_L$ is extracted. Our calculation shows that the Lyapunov exponent monotonically increases as the Kondo coupling $J_K$ increases, and it can reach an order of $\lambda_L\sim T$ as $J_K$ approaches the $MCK$ point. Furthermore, we also demonstrate that $\lambda_L$ decreases monotonously as the impurity and bosonic bath coupling $g$ increases, which is contrary to the general expectation that the most chaotic property occurs at the quantum critical point with the non-Fermi liquid nature.
\end{abstract}
 \maketitle

\section{Introduction}
Recently, the study on quantum chaos in the many-body physics has drawn intensive interests\cite{Swingle2013,Sachdev2015,Chowdhury2017,Igor2016,Song2017,Sumilian2017,Sachdev2018RMP,B.Swingle2018,Han2020}. Quantum chaos can be diagnosed by the so-called out-of-time-ordered correlators (OTOCs)\cite{Shenker2014,Kitaev2014}. The behaviors of OTOCs have been investigated both theoretically and experimentally\cite{Rigol,Langen,Kaufman,zhu2016,Yao2016,Bentsen2019}. The OTOC was first introduced in the context of  superconductivity \cite{Larkin1969} and then generalized to study the information scrambling in the black hole close to the horizon. Usually, it is convenient to define the ``regulated" OTOC\cite{Chowdhury2017,klug2018,sachdevpnas2017,Yao2018},
\bea
\mathcal C(t)={\rm Tr}\{\sqrt \rho [\hat{W}(t), \hat{V}(0)]^\dagger \sqrt \rho[\hat{W}(t), \hat{V}(0)]\}.
\eea
here $\rho=e^{-\beta H}$ is the thermal density at the temperature $T=1/\beta$. And $\hat{W}$ and $\hat{V}$ are local operators. In a chaotic system the OTOC is expected to have an exponential growth $C(t)\propto e^{\lambda_L t}$ at the intermediate time, where $\lambda_L$ is called Lyapunov exponent. Given a perturbation the initial quantum entanglement will spread across all system after a typical time scale $t_{sr}\sim \lambda_L^{-1}$. During this process the initial information is lost and the system goes into the state of thermalization. Under some reasonable conditions, the Lyapunov exponent $\lambda_L$ is proven to have an upper bound $\lambda_L\le 2\pi k_B T/\hbar$\cite{Maldacena2016} and saturates in the models with gravity duals. The most-celebrated $0+1D$ SYK${_4}$ model with random all to all  interactions\cite{Sachdev1993,Kitaev,Maldecena} is a concrete example.

The Kondo model describes the systems in which the impurity spin strongly interacts with conduction electrons. Recently, the information scrambling in the two-channel and one-channel Kondo model has been investigated by mapping them onto the Majorana resonant level models\cite{Dora}. Their results show that the OTOC for the impurity spin in two channel Kondo model is temperature independent and saturates to $1/4$ at late time, while the OTOC in one channel Kondo model vanishes at late time, indicating the absence of the butterfly effect. The Bose-Fermi Kondo model (BFKM) in which the impurity spin interacts with both conduction electron and bosonic bath has rich physical properties, specially the non-Fermi liquid state. From RG analysis\cite{Parcollet1998,Zhu2002}, it contains several nontrivial fixed points. At low temperature and energy limit, a conformal symmetry can emerge at some fixed points. This model is an important system to study the dual of the gravity and many-body physics. Based on non-Fermi liquid behavior and the emergent conformal symmetry, one may expect there are highly chaotic behaviors in this model.

In this paper, we calculate the Lyapunov exponent in the BFKM in the large-N limit. Our calculation shows that there are three types diagrams which have the most important contributions to the OTOCs: two one-rung and one two-rung ladder diagrams. In order to extract the Lyapunov exponent $\lambda_L$ from the Bethe-Salpeter equation, the equal-spaced discretization in energy domain has to been taken. As a consequence of the shortcoming of the method, we can only study quantum chaos at the intermediate temperature region instead of the low temperature region $T\ll T_K^0$ where $T_K^0$ is the bare Kondo temperature.  At the intermediate temperature region we find that $\lambda_T$ decreases monotonically as increasing temperature and the chaotic property will lost at high temperature for given $J_K$. Moreover, for fixed $T$ and $J_K$, our numerical results show the chaotic properties are lost as the coupling between impurity and bosonic bath $g$ increases, violating the expectation that the most chaotic behavior occurs at the quantum critical point with the non-Fermi liquid nature.

The paper is organized as follows. In section \ref{review}, we introduce the BFKM in the large-N limit. In section \ref{Realtime}, we use the Keldysh method to derive the self-consistent equations for Green's functions and solve it with the help of fast-Fourier transformation method. In section \ref{OTOC} the OTOCs are calculated based on the Bethe-Salpter equation in the large-N limit and the Lyapunov exponent is extracted from the OTOCs numerically. In section \ref{results} we demonstrate the results. Finally we summarize the results and give conclusions in section \ref{conclusion}.

\section{Review of the Bose-Fermi Kondo Model}\label{review}
We will start from a system with N fermion flavors. The Hamiltonian can be cast as
\bea
\hat H=&&\sum_{k,\sigma,\alpha}E_{k}c_{k\sigma\alpha}^{\dagger}c_{k\sigma\alpha}+\sum_{k}\epsilon_{k} \Phi_{k}^{\dagger}\Phi_{k}\cr &&+\frac{J_K}{N}\sum_{\alpha=1}^{M}{\bf S} \cdot {\bf s}_{\alpha}+ \frac{g}{\sqrt{N}}\sum_{k}{\bf S} \cdot \big(\Phi_{k}+\Phi^{\dagger}_k\big),
\eea
where $c_{k\sigma\alpha}^{\dagger} (c_{k\sigma\alpha})$ is the creation (annihilation) operator of the conduction electron with channel index $\alpha=1,\cdots, M$ and spin $\sigma=1,\cdots,N$. The conduction electrons at the impurity site transform under the fundamental representation ${\bf s}^{i}_{\alpha}=\sum_{k\sigma\sigma^{\prime}} c_{k\sigma\alpha}^{\dagger}{\bf t}^i_{\sigma\sigma^{\prime}}c_{k\sigma^{\prime}\alpha}(i=1,\cdots,N^2-1)$ and couple to the impurity spin with interaction strength $J_K$. It is convenient to rewrite the impurity spin with $N$ components pseudo-fermion $f_{\alpha}$ as ${\bf S}^i=\sum_{\sigma\sigma^{\prime}} f^{\dagger}_{\sigma} S^i_{\alpha\alpha^{\prime}} f_{\sigma^{\prime}}$ by taking antisymmetric representation\cite{Parcollet1998} with constraint $\sum_{\sigma} f^{\dagger}_{\sigma} f_{\sigma}=Q$ which can be absorbed into action by introducing the Lagrange multiplier $\mu$. In this paper, we consider the case with particle-hole symmetry, requiring $Q=N/2$. $g$ is the interaction strength with bosonic bath $\Phi_k$ with $N^2-1$ independent components which comes from the spin or magnetic fluctuation. The ratio between $M$ number and $N$ is denoted as $\kappa$, which is taken as $\kappa=1/2$ in the following of the paper. The density of state  of conduction electron $A_{c}(\omega)$ around Fermi surface can be approximately as $A_{c}(\omega)=\sum_{k}\delta(\omega-E_k)=\rho_0$ for $|\omega|<D/2$, where $D=1/\rho_0$ is the band width. The Greens' function for boson bath in the imaginary time is denoted as $\mathcal{G}_{\Phi}(\tau)\equiv -\langle \mathcal{T}\Phi(\tau) \Phi^{\dagger}(0)\rangle$. The bosonic spectrum $A_{\Phi}(\omega)=-\Im \mathcal{G}_{\Phi}(\omega+i\eta)/\pi=\sum_{k}[\delta(\omega-\epsilon_k)-\delta(\omega+\epsilon_k)]$ is considered as the sub-Ohmic bosonic spectrum, namely
\bea
A_{\Phi}(\omega)=|\omega|^{1-\epsilon} \text{sign}(\omega),
\eea
for $|\omega|<\Lambda$. And the parameter $\epsilon$ is at the range $[0,1)$. In this  paper we set $\hbar=1$ and $k_B=1$.

\begin{figure}[t]
\includegraphics[width=0.45\textwidth]{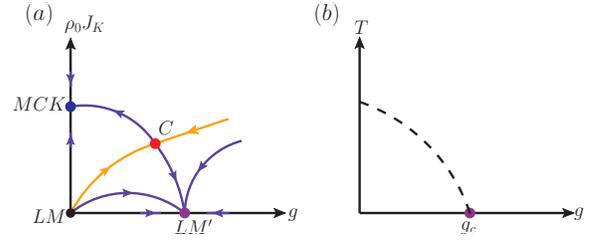}
\caption{(Color online) The general RG flows and the phase transitions in the BFKM. (a) RG flows\cite{Parcollet1998,Zhu2002} in the parameter space $(g,\rho_0 J_K)$. Obviously, there exist three nontrivial fixed points, one unstable quantum critical fixed point $C$ marked by the red point and an critical local moment fixed point $LM^{\prime}$ labeled by the purple point, and the last one $MCK$ is related to the overscreened Kondo phase. The orange line is the line separating Kondo-singlet phase and the disorder phase. (b) The general phase diagram to describe the quantum phase transition. Here $g_c$ denotes the quantum critical point.}
\label{fig:RGphase}
\end{figure}

For the pure multichannel Kondo model, the previous study\cite{Parcollet1998} proves there exists a nontrivial intermediate fixed point $MCK$ between the trivial local moment fixed point $LM$ and strong coupling limit $J_K\rightarrow \infty$ with conformal symmetry. When coupling to the bosonic bath, the another two fixed points can appear\cite{Zhu2002}. One of them is the critical local moment point $LM^{\prime}$ and another one is the unstable critical fixed point $C$. The difference of RG flows between BFKM and Kondo model leads to different chaotic behaviors between them.

\begin{figure}[h]
\includegraphics[width=0.4\textwidth]{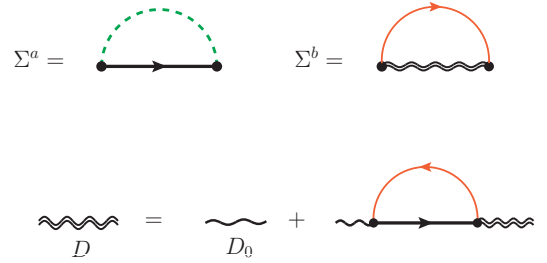}
\caption{(Color online) Feynman diagrams in the large-N limit. The black line represents the full impurity Green's function. Green dashed line is the bosonic bath propagator and red line denotes the propagator of the conduction electrons.  The upper two diagrams are the self-energy corrections for the impurity fermions.}
\label{fig:feynman}
\end{figure}

\section{Green's functions in real time}\label{Realtime}

In order to derive the Green's functions in the real time, it is convinent to rewrite the model in the Keldysh time contour with backward and forward time evolution\cite{Keldysh1965,Kamenev2011}, which is denoted by the sign ``-" and ``+" respectively. Then field $\psi$ regardless of  boson or fermion can be splited into two parts as $\bar{\psi}(t)=\left(\bar{\psi}^{+},\bar{\psi}^{-}\right)$ based on its causal position. After performing Keldysh rotation\cite{Kamenev2011}, the Green's functions are written as

\bea
&&{\bf G}_F=\left(\begin{matrix} G_F^{R}  && G_F^{K}  \cr
0  &&  G_F^{A}
\end{matrix}\right),
 {\bf G}_B(t)=\left(\begin{matrix} G_B^{K}  && G_F^{R} \cr
G_B^{A}  && 0 \end{matrix}
\right),
\eea
where $R$ ($A$) represents the retarded (advanced) Green's function respectively. The Keldysh part, which is denoted by $K$, is related to the retarded and advanced part by the fluctuation-dissipation theorem in the frequency domain,
\bea
&&G_F^K(\omega)=\tanh(\frac{\omega}{2T})\big(G_F^{R}(\omega)-G_F^{A}(\omega)\big),\\
&&G_B^K(\omega)=\coth(\frac{\omega}{2T})\big(G_B^{R}(\omega)-G_B^{A}(\omega)\big).
\eea
Therefore, the noninteracting action can be written as
\bea
&&S_{0}=\int d\omega\Big\{\sum_{\sigma\alpha}\sum_{{\bf k}}\Big(\bar{\Psi}_{\sigma\alpha}(\omega,{\bf k}) {\bf G}^{-1}_{c}(\omega,E_{\bf k})\Psi_{\sigma\alpha}(\omega,{\bf k})+\nonumber \\
&&\sum_{\sigma^{\prime}}\bar{\Phi}_{\sigma\sigma^{\prime}}(\omega,{\bf k})D_{\Phi}^{-1}(\omega,\epsilon_{\bf k}) \Phi_{\sigma\sigma^{\prime}}\Big)+\sum_{\sigma}\bar{\mathcal{F}}_{\sigma}(\omega){\bf G}^{-1}_{0}\mathcal{F}_{\sigma}(\omega)\Big\},\nonumber \\
\eea
where the fields is written in Keldysh space as $\bar{\Psi}_{\sigma\alpha}({\bf k})=\left(c^{\dagger}_{1,{\bf k}\sigma\alpha},c^{\dagger}_{2,{\bf k}\sigma\alpha}\right)$, $\bar{\Phi}({\bf k})=\left(\Phi^{\dagger}_{1,{\bf k}},\Phi^{\dagger}_{2,{\bf k}}\right)$ and $\bar{\mathcal{F}}_{\sigma}=\left(f^{\dagger}_{1\sigma},f^{\dagger}_{2\sigma}\right)$. The Green's function for conduction electron and bosonic bath are given by
\bea
&&G^{R}_c(\omega,{\bf k})=\big(G^{A}_c(\omega,{\bf k})\big)^*=\frac{1}{\omega+i\eta-E_{\bf k}},\\
&&D_{\Phi}^{R}(\omega,{\bf k})=\big(D_{\Phi}^{A}(\omega,{\bf k})\big)^*=\sum_{s=\pm1}\frac{s}{\omega+i\eta-s\epsilon_{\bf k}},
\eea
and the bare Green's function for impurity is given by
\bea
G_0^{R}(\omega)=\frac{1}{\omega+i\eta-\lambda},
\eea
where $\lambda$ is the saddle point of the auxiliary field $\mu$ to force the conservation of impurity electrons and it can be taken as zero because of the particle-hole symmetry $f_{\sigma}\leftrightarrow f_{\sigma}^{\dagger}$. For the interaction between the bosonic bath and impurity, the action is
\bea
&&S_{\Phi,\mathcal{F}}=\frac{g}{\sqrt{2N}}\sum_{\sigma,\sigma^{\prime}}\int dt \Big\{\bar{\mathcal{F}}_{\sigma}(t)\gamma_{1}\mathcal{F}_{\sigma^{\prime}}(t)(\Phi_{1,\sigma\sigma^{\prime}}+\bar{\Phi}_{1,\sigma\sigma^{\prime}})\nonumber \\
&&+\bar{\mathcal{F}}_{\sigma}(t)\gamma_{2}\mathcal{F}_{\sigma^{\prime}}(t)(\Phi_{2,\sigma\sigma^{\prime}}+\bar{\Phi}_{2,\sigma\sigma^{\prime}})\Big\}.
\eea
For interaction between conduction electron and impurity, we introduce $M$-flavors Hubbard-Stratonovich fields $B_{\alpha}$, which leads to
\bea
&&S_{\Psi,\mathcal{B},\mathcal{F}}=\frac{1}{\sqrt{2N}}\sum_{\sigma,\alpha}\int dt \Big\{\bar{\mathcal{F}}_{\sigma}\gamma_{1}\Psi_{\sigma\alpha}B_{1\alpha}+ \bar{\mathcal{F}}_{\sigma}\gamma_{2}\Psi_{\sigma\alpha}B_{2\alpha} \nonumber \\
&&+H.C.\Big\}+\sum_{\alpha}\int d\omega \bar{\mathcal{B}}_{\alpha}{\bf D}_{0}^{-1}(\omega)\mathcal{B}_{\alpha},
\eea
where $\bar{\mathcal{B}}_{\alpha}=\left(B^{\dagger}_{1\alpha},B^{\dagger}_{2\alpha}\right)$. The matrix $\gamma_{1}$ and $\gamma_2$ are given as following,
\bea
\gamma_1=\left(\begin{matrix} 1  && 0  \cr
0  &&  1
\end{matrix}\right), \quad \gamma_2=\left(\begin{matrix} 0  && 1  \cr
1  &&  0
\end{matrix}\right),
\eea
and the bare propagator $D_0$ for $B_{\alpha}$ is
\bea
D_{0}^{R}(\omega)=\big(D_{0}^{A}(\omega)\big)^{*}=-\frac{1}{J_K}.
\eea
Therefore, the partition function is $\mathcal{Z}=\int\mathcal{D}[\Phi,\mathcal{F},\mathcal{B},\lambda] e^{i\big(S_0+S_{\Phi,\mathcal{F}}+S_{\Psi,\mathcal{B},\mathcal{F}}\big)}$. Due to interaction, the Green's functions will be renormalized and the self-energy for fermions or bosons has the following structure,
\bea
{\bf \Sigma}_F=\left(\begin{matrix} \Sigma_F^{R}  && \Sigma_F^{K}  \cr
0  &&  \Sigma_F^{A}
\end{matrix}\right), \quad
{\bf \Sigma}_B=\left(\begin{matrix} \Sigma_B^{K}  && \Sigma_F^{R} \cr
\Sigma_B^{A} && 0 \end{matrix}
\right).
\eea
\begin{figure}[ht]
\includegraphics[width=0.48\textwidth]{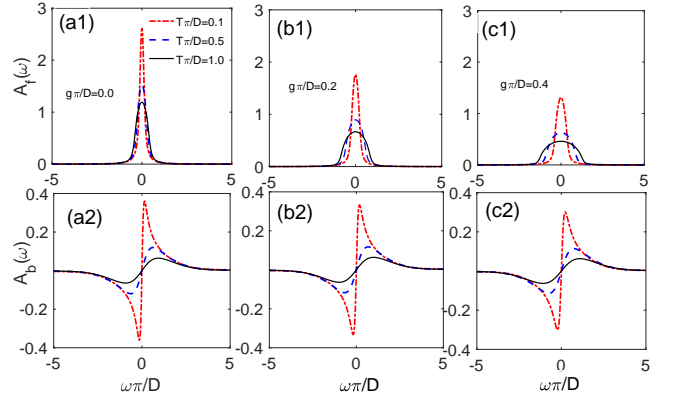}
\caption{(Color online) The numerical results for the impurity spectral function $A_f(\omega)=-\Im G^R(\omega)/\pi$ and bosonic spectral functions $A_b(\omega)=-\Im D^R(\omega)/\pi$ at the fixed Kondo coulpling $J_K\pi/D=1.0$ and at the sub-ohmic case $\epsilon=0.5$. The red dash-dotted, blue dashed, and black solid lines correspond to the results at temperature $T\pi/D=0.1$, $0.5$ and $1.0$ respectively. (a1,a2), (b1,b2), and (c1,c2) are the spectral functions at the bosonic bath interaction $g\pi/D=0$, $0.2$ and $0.4$ respectively.}
\label{fig:spectral}
\end{figure}

The self-energies are obtained by taking into account the most relevant Feynman diagrams in the large-N limit, shown in Fig.~\ref{fig:feynman}. Therefore it is straightforward to obtain the following self-consistent equations,
\bea\label{Eq:MainEq}
&&\big(G^{R}(\omega)\big)^{-1}=\omega-\lambda-\Sigma_{a}^R(\omega)-\Sigma_{b}^R(\omega),\\
&&i\Sigma_{a}^R(t)=\frac{g^2}{2}\Big( D_{\Phi}^{K}(t) G^R(t)+D_{\Phi}^R(t) G^{K}(t)\Big), \\
&&i\Sigma_{b}^R(t)=\frac{\kappa}{2}\Big( D^{K}(t) G_{c}^R(t)+ D^R(t) G_c^{K}(t)\Big),\\
&&\big(D^{R}(\omega)\big)^{-1}=-1/J_K-\Pi^{R}(\omega), \\
&&i\Pi^{R}(t)=\frac{1}{2}\Big(- G^{K}(-t) G_{c}^R(t)-G^A(-t) G_c^{K}(t)\Big).
\eea
Here $D_{\Phi}^{R}(t)=\int \frac{d\omega}{2\pi} D^{R}_{\Phi}(\omega)e^{i\omega t}$ and $G^{R}_c(t)=\int \frac{d\omega}{2\pi} G_c^{R}(\omega)e^{i\omega t}$. The Green's function for conduction electron and bosonic bath is obtained by using the Kramers-Kronig relation: $G_{c}^{R}(\omega)=\int d\omega^{\prime}\frac{A_{c}(\omega^{\prime})}{\omega+i\eta-\omega^{\prime}}$ and $D_{\Phi}^{R}(\omega)=\int d\omega^{\prime}\frac{A_{\Phi}(\omega^{\prime})}{\omega+i\eta-\omega^{\prime}}$.

To obtain the impurity and bosonic Green's functions, we numerically solve the self-consistent equations by the fast-Fourier transformation (FFT) method. In practice, the electron spectral is taken as a Gaussian function $A_c(\omega)=e^{-\omega^2/\pi}/\pi$. In the Fig. \ref{fig:spectral}, we plot the impurity and bosonic spectral functions respectively for different temperatures $T$ and bosonic coupling $g$ while fixing the Kondo coupling $J
_K\pi/D=1.0$. From Fig. \ref{fig:spectral}, \begin{figure}[h]
\includegraphics[width=0.4\textwidth]{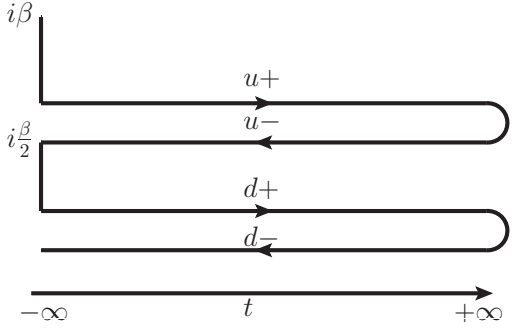}
\caption{The augmented Keldysh time contour for calculating the out-of-time-oredered correlators. The horizontal direction represents the real time evolution and the vertical direction represents the imaginary time evolution. It contains two Keldysh time contours, which are separated by $i\beta/2$ and labeled by $u$ and $d$ respectively. Each Keldysh time contour contains two real time evolutions, the forward one and the backward one.}
\label{fig:timecontour}
\end{figure}one can observe amplitudes of impurity and bosonic spectral functions both decrease as increasing the bosonic bath coupling $g$.

\section{Out of time correlator and Bethe-Salpeter Equation}\label{OTOC}

It is convenient to evaluate the retarded ``regulated" squared anti-commutator defined as \cite{Chowdhury2017,klug2018,Yao2018}
\bea
\mathcal C(t_1,t_2)=&&\frac{\theta(t_{1})\theta(t_{2})}{N^2}\sum_{\sigma,\sigma^{\prime}}{\rm Tr}\Big(\sqrt{\rho}\{f_{\sigma}(t_1),f_{\sigma^{\prime}}^{\dagger}(0)\} \cr &&
\times\sqrt{\rho}\{f_{\sigma}(t_2),f_{\sigma^{\prime}}^{\dagger}(0)\}^{\dagger}\Big),
\eea
where $\rho=\exp(-\beta H)$ is the thermal density matrix. It is clear that the OTO is defined in the two-copied Keldysh contours\cite{Igor2016} separated by the imaginary time $i\beta/2$ as shown in Fig. \ref{fig:timecontour}. Here we denote each Keldysh contour with indices $s=(u,d)$. Therefore the fermionic or bosonic field $\psi$ in the two-copied Keldysh contours is generalized to $\bar{\psi}=\left(\psi^{\dagger}_{u,cl},\psi^{\dagger}_{u,q},\psi^{\dagger}_{d,cl},\psi^{\dagger}_{d,q}\right)$ after performing the Keldysh rotation for each time fold. Moreover the Green's function in each time fold remains the same, while the interloop Green's function ${\bf G}_{s\bar{s}}$ or ${\bf D}_{s\bar{s}}$ ($\bar{d}=u$ and $\bar{u}=d$) has the following structure,
\bea
&&{\bf G}_{s\bar{s}}=\left(\begin{matrix} 0  && G_{s\bar{s}}^{K}  \cr
0  &&  0
\end{matrix}\right),
{\bf D}_{s\bar{s}}(t)=\left(\begin{matrix} D_{s\bar{s}}^{K}  && 0 \cr
0  && 0 \end{matrix}
\right),
\eea
here the component $G_{s\bar{s}}^{K}$ or $D_{s\bar{s}}^{K}$ has the generalized fluctuation-dissipation theorem,
\bea
&&G_{ud}^K(\omega)=\big(G_{du}^K(\omega)\big)^*=\frac{2i \Im G^{R}(\omega)}{\cosh(\frac{\omega}{2T})},\\
&&D_{ud}^K(\omega)=D_{du}^K(\omega)=\frac{2i \Im D^{R}(\omega)}{\sinh(\frac{\omega}{2T})}.
\eea

\begin{figure}[ht]
\includegraphics[width=0.5\textwidth]{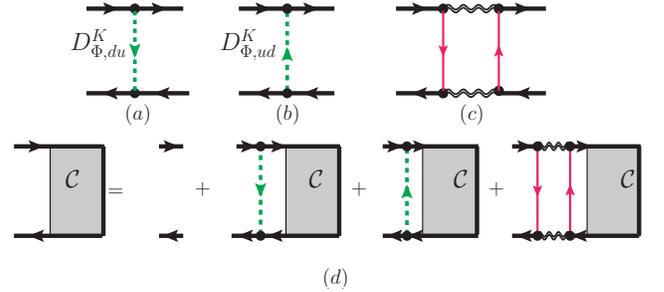}
\caption{(Color online) Diagrammatically representations of the Bethe-Salpeter equations.}
\label{fig:BS}
\end{figure}

In the augmented Keldysh space, the OTO correlator $\mathcal{C}$ can be rewritten as
\bea
&&\mathcal{C}(t_1,t_2)=\nonumber \\
&&-\frac{\theta(t_1)\theta(t_2)}{N^2}\sum_{\sigma\sigma^{\prime}}\Big\langle f^{d,cl}_{\sigma}(t_1) \bar{f}^{d,cl}_{\sigma^{\prime}}(0) \bar{f}^{u,q}_{\sigma^{\prime}}(0) f^{u,q}_{\sigma}(t_2)\Big\rangle_{aK}
\eea

where $\langle \dots\rangle_{aK}=\int \mathcal{D}[\Phi,\mathcal{F},\mathcal{B},\lambda]e^{iS_{aK}}$ is the average in the augmented Keldysh contours. In large-N limit, vertex correction is ignored, hence we use bare vertex in Bethe-Salpeter equation as illustrated in Fig. \ref{fig:BS} to evaluate the quantum chaos in OTOC. The diagram contains two types of ladder diagrams, the first type are the two one-rung diagrams with $\Phi$ field connecting the upper and down worlds, and the second type is the two-rungs diagram with conduction electron fields linking the different worlds. Here are Feynman rules for these diagrams: (i) the rail lines in the upper world represents the advanced Green's functions; (ii) the rail lines sited in the down world are retarded Green's functions; (iii) the rungs connecting two worlds corresponds to the $G^{K}_{ud(du)}$ or $D^{K}_{ud(du)}$.

Since there is no dissipation, the time translation symmetry holds. Hence we can take the following Fourier transformation,
\bea
\mathcal{C}(t_1,t_2)=\frac{1}{N}\int \frac{d\Omega d\omega}{(2\pi)^2} e^{-i\Omega(t_1-t_2)-i\omega t} \mathcal{C}(\Omega,\omega),
\eea
here we introduce the center of mass time separation $t=\big(t_1+t_2\big)/2$. Following the aforementioned rules to calculate the OTOC, the zero order for $\mathcal{C}(\Omega,\omega)$ is
\bea
\mathcal{C}_0(\Omega,\omega)=G^{R}(\Omega+\frac{\omega}{2}) G^A(\Omega-\frac{\omega}{2})\equiv \mathcal{A_{\omega}}(\Omega),
\eea
and $\mathcal{A}_{\omega}(\Omega)$ is a positive real number due to $G^{R}(\Omega+\omega/2)=\big(G^A(\Omega-\omega/2)\big)^*$. Summing up the ladder diagrams, we can obtain the Bethe-Salpeter equation,
\bea\label{eq:BSequation}
&&\mathcal{C}(\Omega,\omega)=\mathcal{A}_{\omega}(\Omega)\Big\{1+\int\frac{d\Omega^{\prime}}{2\pi}\Big(\mathcal{K}_{1,\omega}(\Omega,\Omega^{\prime})+\mathcal{K}_{2,\omega}(\Omega,\Omega^{\prime})\Big)\nonumber \\
\times &&\mathcal{C}(\Omega^{\prime},\omega)\Big\},
\eea
followed by one-rung kernel $\mathcal{K}_{1,\omega}$,
\bea
\mathcal{K}_{1,\omega}=\frac{ig^2}{2}D_{\Phi,ud}^{K}(\Omega^{\prime}-\Omega)+\frac{ig^2}{2}D_{\Phi,du}^{K}(\Omega-\Omega^{\prime}),
\eea
and one two-rungs kernel $\mathcal{K}_{2,\omega}$,
\bea \label{Eq:Kernel}
&&\mathcal{K}_{2,\omega}(\Omega,\Omega^{\prime})=\frac{\kappa}{4} \int \frac{d\omega^{\prime}}{2\pi}\Big(D^{R}(\omega^{\prime}+\frac{\omega}{2}) \nonumber \\&&
\times G^{K}_{c,ud}(\Omega-\omega^{\prime}) G^{K}_{c,du}(\Omega^{\prime}-\omega^{\prime})D^{A}(\omega^{\prime}-\frac{\omega}{2})\Big).
\eea
Finally, one can get the following irreducible ladder diagrams after dropping the irrelevant inhomogeneous term in Eq. \ref{eq:BSequation},
\bea\label{Eq:BSeq}
\mathcal{C}(\Omega,\omega)=\mathcal{A}_{\omega}(\Omega)\int\frac{d\Omega^{\prime}}{2\pi}\bar{\mathcal{K}}_{\omega}(\Omega,\Omega^{\prime}) \mathcal{C}(\Omega^{\prime},\omega).
\eea
Here $\bar{\mathcal{K}}_{\omega}\equiv \mathcal{K}_{1,\omega}+\mathcal{K}_{2,\omega}$. We clarify that the above equation we obtained contains leading order contributions in the large-N limit. While for finite N, the vertex correction might be relevant expecially for strong coupling and one need take into account higher order diagrams. The Lyapunov exponent $\lambda_L$ corresponds to the positive solution of $-i\omega$ to make the integral kernel in the Bethe-Salpeter equation have an unite eigenvalue\cite{sachdevpnas2017,Gu2019}. The existence of positive solution will signal the chaotic behavior, and on the other hand, it implies the absence of the chaotic properties. The details of the numerical calculation of Lyapunov exponent can be found in Appendix. \ref{NC:Lyapunov}.

\section{Numerical Results}\label{results}
In this section, we give the numerical results of the Lyapunov exponent at the $\mathcal{O}(1)$ order at the intermediate temperature in the large $N$ limit. From the Eq. \ref{Eq:Kernel} and Eq. \ref{Eq:BSeq}, it can be found that the eigenvalue $\lambda_i$ of the kernel matrix $\mathcal{A}_{\omega}\mathcal{K}_{\omega}$ at $g=0$ is proportional to the square of Kondo coupling $J_K$ at the weak coupling limit, $\lambda_i\propto J_K^2$. Hence the condition for the existence of unit eigenvalue for given $-i\omega$ cannot be satisfied because of $\lambda_i\propto J_K^2\ll 1$, implying there is no chaotic behavior at the weak coupling limit for the pure multichannel Kondo model. This observation can be conformed by the following numerical results.

\begin{figure}[h]
\includegraphics[width=0.4\textwidth]{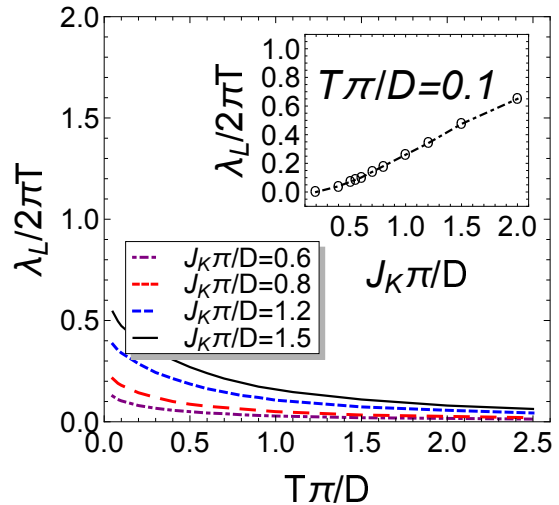}
\caption{(Color online) $\lambda_L  /2 \pi T$ as a function of temperature $T\pi/ D$ for the BKFM at $g\pi/D=0$. The black solid, the blue dashed, the red dashing and the purple dot-dashed curves correspond to different Kondo coupling $J_K\pi/D=1.5$, $1.2$, $0.8$ and $0.6$. Insert: $\lambda_L/2\pi T$ as a function of $J_K\pi/D$ for fixed temperature $T\pi/D=0.1$.}
\label{fig:NR}
\end{figure}

In Fig. \ref{fig:NR} we plot the ratio $\lambda_L/2\pi T$ as a function of temperature $T$ at the fixed Kondo coupling $J_K\pi/D=0.6$, $J_K\pi/D=0.8$, $J_K\pi/D=1.2$ and $J_K\pi/D=1.5$ in absence of coupling to bosonic bath $g=0$. For a fixed $J_k$, the Lyapunov exponent always decrease monotonically as increasing temperature. Furthermore, the chaotic behavior do not exist anymore when the temperature reaches to a critical value, say $T^*$. One can expect this result because of its transition from non-Fermi liquid character to the Fermi liquid nature during the process of increasing temperature $T$. When the Kondo coupling $J_K$ growths from the local moment fixed point ($LM$), it will reach the nontrivial overscreen multichannel fixed point ($MCK$) where the conformal symmetry will emergent\cite{Parcollet1998}. One can also observe the ratio $\lambda_L/T$ growths monotonically as reaching to the $MCK$ point as shown in the Fig. \ref{fig:NR}, specially shown in the inset figure in the Fig. \ref{fig:NR} at $T\pi/D=0.1$. The $\lambda_L=0$ at $J_K\pi/D\approx0.21$ confirm our aforementioned analysis that the chaotic behavior vanish at the weak Kondo coupling limit $J_K\ll1$. One should note that the statement is obtained for finite temperature and is not neccessary true for zero temperature when scaling relation between Green's function develops (see the appendix C).

\begin{figure}[h]
\includegraphics[width=0.48\textwidth]{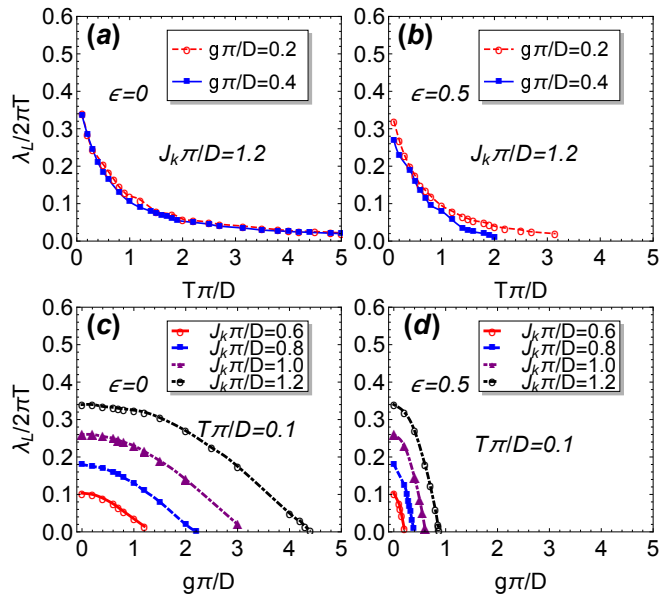}
\caption{(Color online) The ratio $\lambda_L/2\pi T$ as a function of temperature $T\pi/D$ for (a,b) and as a function of boson coupling $g$ for (c,d).}
\label{fig:gneq0}\label{Fig:evolutionforg}
\end{figure}

Now we introduce the coupling between impurity and bosonic bath $g$. By increasing $g$, the systems will go through the overscreened multichannel Kondo phase to a critical local moment phase which is separated by a unstable fixed point\cite{Zhu2002,Zhu2004,Kirchner,Zuodong} with critical coupling $g_c$. Generally near the quantum critical point, the non-Fermi liquid will arise and the conformal symmetry emerges, leading to the larger chaotic behavior with $\lambda_L\sim T$ than other regions away from critical point\cite{Shen2017}.  To check this argument, we plot the $\lambda_L/2\pi T$ as a functions of temperature at different $g\pi/D$ and as functions of $g\pi/D$  at a fixed temperature $T\pi/D=0.1$ as illustrated in Fig. \ref{Fig:evolutionforg}.  In Fig. \ref{Fig:evolutionforg} (a,b) we can obtain the two main observations: (1) For sub-ohmic case $\epsilon=0.5$, the Lyapunov exponent decreases as growing bosonic coupling $g$ at the same parameters $(J_K,T)$. It  shares the same behavior to the ohmic case $\epsilon=0$ although they have quiet different RG flows. This fact indicate the violation of the above argument. (2) Butterfly effect is stronger at the ohmic case than the sub-ohmic one, by comparing Fig. \ref{Fig:evolutionforg} (a) to  Fig. \ref{Fig:evolutionforg} (b). In order to investigate the behavior of Lyapunov exponent when crossing critical point, we performed a detailed $g$ dependence calculation for various $J_{k}$ at fixed temperature as shown in Fig. \ref{Fig:evolutionforg} (c,d). One can clearly see the Lyapunov exponent is indeed monotonically decreasing as increasing $g$, and is finally vanishing at finite $g$. This monotonous behavior is consistent with the result that residual entropy for this model increases monotonously from $MCK$ phase to $LM^{\prime}$ phase\cite{Zuodong}. This behavior of residual entropy voilate the g-theorem which demand the entropy should decrease along RG trajectories if conformal invariance is presented. As the original model accually break conformal symmetry in sub-ohmic case, one may not expect the largest chaotic behavior at critical point similar to impurity entropy case. From numerical perspective, the reason of the decreasing of $\lambda_L$ with $g$ can be attribute to the suppressive effect of magnitude of spectral functions of impurity and auxiliary boson as shown in Fig. \ref{fig:spectral}.

\section{Conclusions}\label{conclusion}
In this paper, we derive the Bethe-Salpeter equation for our defined impurity OTO correlator for BFKM in large-N limit. We find the biggest contribution comes from two one-rung and one two-rungs diagram whose upper and down world lines are connected by bosonic bath and the conduction electrons respectively. The numerical calculation at the intermediate temperature shows that the Lyapunov exponent $\lambda_L$ decreases with increasing temperature and finally vanishe at a finite temperature, which is different with the one and two-channel Kondo models in which impurity OTOC is temperature-independent. We also observe the system has no butterfly effect below a typical Kondo coupling $J_K$ for finite temperature. When coupled to bosonic bath, the monotonously decrease of $\lambda_L$ do not obey general argument that the highest chaotic behavior occurs at the quantum critical point, but is consistent with the behavior of impurity entropy and violation of g-theorem in the model\cite{Zuodong}.

\section{Acknowledgments}
We thank Boyang Liu and Pengfei Zhang for very helpful discussions. The work is supported by the Hong Kong Research Grants Council, GRF 17304719, CRF C6026-16W and C6005-17G.

\appendix
\section{Numerical technique to solve the saddle point equations}\label{NC:Green}

The Fourier transformation method is used to solve the self-consistent equations \ref{Eq:MainEq} and also the integral kernel $\mathcal{A}_{\omega}\mathcal{K}_{\omega}$ in the main text. We discrete the frequency and time domain as
\bea
&&\Omega_{d}=\frac{2\pi f_s}{N_t}[\frac{1-N_t}{2},\frac{3-N_t}{2},\dots,\frac{N_t-3}{2},\frac{N_t-1}{2}],\\
&&T_{d}=\frac{1}{f_s}[\frac{1-N_t}{2},\frac{3-N_t}{2},\dots,\frac{N_t-3}{2},\frac{N_t-1}{2}].
\eea
The Fourier transformation $G(t)=\int \frac{d\omega}{2\pi}G(\omega)e^{-i\omega t}$ and $G(\omega)=\int_0^{+\infty} dt G(\omega) e^{i\omega t}$ for Green's functions can be performed by the FFT algorithm. We iteratively solve the equations and obtain the self-consistent solutions if the error $max(|G(\omega)-\bar{G}(\omega)|)$ where $G$ and $\bar{G}$ belong to two nearest iterative steps is less than $10^{-6}$. In practice, the total point $N_t=2^{21}+1$ and $f_s=4$ is used. The cutoff for the frequency is $\omega_{c}=4\pi$ which is is four times of band width $D$. Moreover, in our numerical calculations the spectral function of the bosonic bath is taken as,
\bea
A_{\Phi}(\omega)=\left\{\begin{aligned}& |\omega|^{1-\epsilon}\text{sign}(\omega), |\omega|<\Lambda \\
& |\Lambda|^{1-\epsilon}\text{sign}(\omega) e^{-c(|\omega|-\Lambda)}, |\omega|\ge \Lambda
\end{aligned}\right.
\eea
here energy cutoff $\Lambda=0.05$ and $c=15$.
\begin{figure}[h]\label{Fig:Jk12forg}
\includegraphics[width=0.45\textwidth]{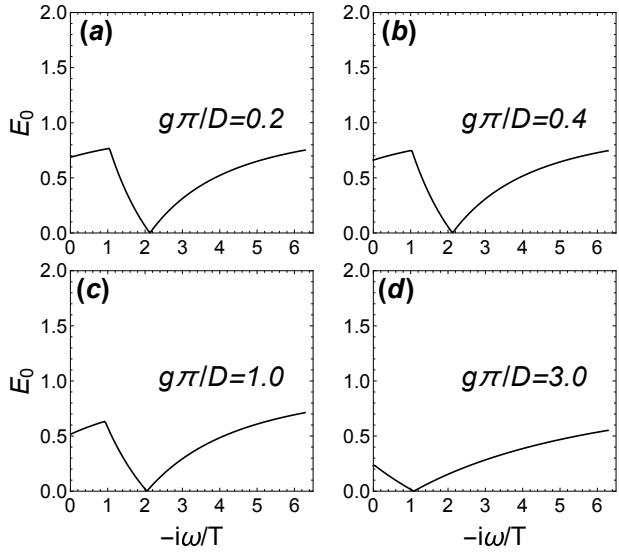}
\caption{(Color online) Plot the magnitude of $E_0$ as a function of $-i\omega/T$ in the positive axis at the given $J_K\pi/D=1.2$ for $g\pi/=0.2$ (a), $0.4$ (b), $1.0$ (c) and $3.0$ (d).}
\label{fig:E0}
\end{figure}
\section{Numerical method for the Lyapunov exponent}\label{NC:Lyapunov}
To numerically calculate the Lyapunov, we first discrete the frequency to transform the integral equation to a linear algebra equation where the integral kernel $\mathcal{A}_{\omega}(\Omega) \mathcal{K}_{\omega}(\Omega,\Omega^{\prime})$ becomes a matrix. The Lyapunov exponent corresponds to the positive value $-i\omega$ which leads to existence of unite eigenvalue of the kernel matrix. Here the energy is discretized in range $(-5,5)$ with total number $N_{size}=800$. The consistency of the result for $N_{size}$ is checked with smaller interval. In Fig. \ref{fig:E0}, we illustrates the evolution of $E_0=\min|1-\lambda_i|$ where $\lambda_i$ is the eigenvalue of the integral kernel of the Bethe-Salpeter equation Eq. \ref{Eq:BSeq}.

\section{Bethe-Salpeter equation at the low temperature limit}
Argument in the main text that $\lambda_i\propto J_K^2$ at the intermediate temperature when absence of bosonic bath cannot be applied the case at the low temperature limit. For $g=0$ case, the flow diagram tells us that the system will flow to $MCK$ point at any finite $J_K$. Then we can apply the scaling ansatz:
\bea
&&\Im G(\omega)=-M_f|\omega|^{\alpha_f-1},\\
&&\Im D(\omega)=-M_B |\omega|^{\alpha_B-1}
\eea
for fixed point Green's functions at zero temperature\cite{Parcollet1998}. Insert these relations into self-consistent equations one can obtain the relation for amplitudes and exponents:
\bea
&&\alpha_f=1-\alpha_B=\frac{1}{1+\kappa}, \\
&&M_c M_f M_B \propto \alpha_f \tan(\frac{\pi\alpha_f}{2})
\eea
for $MCK$ fixed point at $g=0$ case. Where $M_c$ is the amplitudes prefactor for condition electron. This means that at zero temperature, the solution of Bethe-Salpeter Eq. \ref{eq:BSequation} will not dependent on the value of $J_K$, i.e., the eigenvalue $\lambda_i$ will saturate to the same value for any finite $J_K$, similar to the entropy results\cite{Zuodong}.

\end{document}